\documentstyle[epsfig,12pt]{article}
\def\Journal#1#2#3#4{{#1}{\bf #2},#3 (#4)}
\def\prd{{\em Phys. Rev.} D}

\newcommand{\beq}[1]{\begin{equation}\label{#1}}
\newcommand\eeq{\end{equation}}
\newcommand{\ba}[1]{\begin{eqnarray}\label{#1}}
\newcommand{\baa}{\begin{eqnarray}}
\newcommand\ea{\end{eqnarray}}
\newcommand{\bee}{\begin{equation}}

\def\l{\lambda}

\newcommand{\h}{Hamiltonian}

\def\hlf{\frac{1}{2}}

\begin{document}

\begin{center}
{\Large \bf Time operator for a quantum many-body system with
SU(1,1) dynamical symmetry}
\vspace{0.5cm} \\
Ivan Andri\' c and  Mladen Martinis \\
{\em Division of Theoretical Physics \\
Rudjer Bo\v skovi\' c Institute, Zagreb, Croatia} \\
\end{center}
\footnote{e-mail address: iandric@rudjer.irb.hr \\ \hspace*{2.2cm}
martinis@rudjer.irb.hr }
\begin{abstract}
 The time operator canonically  conjugated to the
Hamiltonian of $N$ interacting particles on the line is
constructed using $SU(1,1)$ as a dynamical symmetry. This hidden
conformal symmetry enables us to make a group theoretic analysis
of the time operator in terms of  $SU(1,1)$ generators. At
distances very far from the interacting region the time operator
is represented as a generalization of the quantum time-of-arrival
operator.

\end{abstract}

The question whether time is a quantum mechanical observable or
not can be formally posed as a question if there exits some
operator conjugated to the Hamiltonian \cite{mug}. If the
Hamiltonian describes a complicated quantum system, such as an
$N$-body system with the interaction between  particles, we expect
that the conjugate time operator, having dimension of time, will
be an observable of the system providing additional complementary
information. We shall explicitly construct the time operator for a
given $N$-particle system and show some of its properties.

In this letter we  study a nonrelativistic many-body system in one
dimension with the scale-invariant $1/r^2$ potential \cite{c}.
Such a system possesses  additional invariance: the overall time
can be arbitrarily reparametrized \cite{aff}. This is a symmetry
of the action and not of the Lagrangian itself. These
transformations are: time translation leading to energy
conservation with the Hamiltonian as the generator, time
dilatations, and time special conformal transformations.

We  investigate a system of  $N$ particles on the line interacting
with the potential proportional to the inverse-square distance. A
proper treatment of the model is in the center-of-mass system
owing to translational invariance. If we are not treating in the
CM, the difference is of the order $1/N$ and  is vanishing in the
large-$N$ limit. The Hamiltonian is of the Calogero-Moser \cite{m}
type

\begin{equation}
H_{CM}=\hlf\sum_{i=1}^Np_i^2+\frac{\l(\l-1)}{2}\sum_{i\neq
j}^N\frac{1} {(x_i-x_j)^2},
 \eeq

where $x_i, \; \; i = 1,2,\cdots N$ represents a particle
coordinate and the dimensionless coupling constant $g = \l (\l -
1)$ is parametrized in terms of the statistical parameter $\l $,
and $\hbar, m = 1$. Introducing completely translationally
invariant variables \cite{Pere}

\begin{equation}
\xi_i=x_i-X,\;\;\partial_{\xi_i}\xi_j=\delta_{ij}-\frac{1}{N},\;\;\;
X = \frac{1}{N}\sum_{i=1}^N x_i, \eeq

we can separate the center-of-mass degrees of freedom. The wave
function of the problem (1) will contain the Jastrow factor,
 therefore it is  convenient to perform  a similarity
 transformation of the \h \ into
\begin{equation}
\prod_{i<j}^N(x_i-x_j)^{-\l}(-H_{CM})\prod_{i<j}^N(x_i-x_j)^{\l}=
\hlf\sum_{i=1}^N\partial_i^2+\frac{\l}{2}\sum_{i\neq
j}^N\frac{1}{x_i-x_j} (\partial_i-\partial_j) .
\eeq

Eliminating the center-of-mass degrees of freedom, we obtain the
generator \cite{and}
\begin{equation}
T_+=\hlf\sum_{i=1}^N\partial_{\xi_i}^2+\frac{\l}{2} \sum_{i\neq
j}^N\frac{1}{\xi_i-\xi_j}(\partial_{\xi_i}-\partial_{\xi_j}),
 \eeq

and the generators of  scale and special conformal invariance are,
respectively,

\begin{equation}
T_0 = -\hlf(\sum_{i=1}^N\xi_i\partial_{\xi_i}+E_0-\hlf),\;\; T_- =
\hlf\sum_{i=1}^N\xi_i^2,
\eeq

Using (2), we can verify that
\begin{equation}
[T_+,T_-]=-2T_0,\;\; [T_0,T_{\pm}]=\pm T_{\pm}.
\eeq

This  is the usual SU(1,1) conformal algebra with the Casimir
operator

\begin{equation}
\hat C=T_+T_--T_0(T_0-1).
\eeq

In the definition  of the operator $T_0$ the  constant $E_0$ is
$E_0=\frac{\l}{2}N(N-1)+\frac{N}{2}$ for consistency reasons,
 and $-1/2$ appears after  the center-of-mass degrees
 of freedom are removed This is the  difference between
 our treatment and that in \cite{Gur}.

{\em 1. Coherent state solution.}Using the established
representation of the $su(1,1)$ algebra, we show that the Calogero
solutions are completely determined assuming that the zero-energy
solutions are known \cite{c}:

\begin{equation}
T_+P_m = 0, \;\;\; T_0P_m = \mu_mP_m,
\end{equation}

where $\mu_m = -\hlf(m+E_0-\hlf)$. Calogero has proved that the
zero-energy solutions $P_m$ are scale- and translationally
invariant homogeneous multivariable polynomials of degree $m$,
written in the center-of-mass variables. Let us assume that
nonzero energy states of the operator $T_+$ are obtained by
applying  a function of the generators \cite{and}to the ground
state:

\begin{equation}
\Psi(T_-,T_0,T_+)P_m = \sum_{p,q,n}c_{pqn}T_-^pT_0^qT_+^n P_m
=\Psi_m(T_-)P_m.
\end{equation}

$\Psi_m(T_-)$ will be determined from the eigenvalue equation

\begin{equation}
-T_+\Psi_m(T_-)P_m=E\Psi_m(T_-)P_m.
\eeq

Using Eq. (6) we can derive the formula

\begin{equation}
[T_+,f(T_-)]=T_-f''(T_-)-2f'(T_-)T_0
 \eeq

and from (10) and (11) we obtain  the Calogero solution (in his
notation, $p=\sqrt{2E}$, $r^2=2T_-$)

\begin{equation}
\Psi_{m, E}(T_-)P_m \sim T_-^{(1-m-E_0+1/2)/2}{\cal Z}
_{m+E_0-3/2}(2\sqrt{ET_-})P_m.
\eeq

By expanding the Bessel function $\cal Z$ in (12) with the
repeated use of the formula

\begin{equation}
 T_-^n \frac{1}{-T_0-\mu + n} = \frac{1}{-T_0 - \mu}T_-^n ,
\end{equation}

together with (8), we finally obtain
\begin{equation}
\Psi_{m, E}(T_-)P_m \sim e^{E\hat{T}}P_m ,
\eeq

 where
\begin{equation}
\hat{T} = T_- \frac{-1}{T_0 +\mu_c},
\eeq

and $\mu_c = (-1+\sqrt{1-4\hat{C}})/2$ is the positive root of the
Casimir eigenvalue and  $\hat{C}$ is the Casimir operator
\cite{Gur}.

We have shown that the Calogero solution (8) in the operator form
(14) is a generalized Barut-Girardello coherent state \cite{bg}
where the Hamiltonian ($-T_+$ up to similarity transformation)
plays the role of the annihilation operator and $\hat{T}$ the role
of the creation operator with the commutation relation
\begin{equation}
 [T_+, i\hat{T}] = i.
\eeq

{\em 2. Eigenstates of time operator.} It can be verified that the
eigenstates of $\hat{T}$
\begin{equation}
i\hat{T}\psi _t = t\psi _t \eeq

are of the form
\begin{equation}
\psi _t = e^{\frac{i}{t}T_-} Q , \eeq

where $Q$ satisfies
\begin{equation}
T_0 Q = -(\mu _c +1) Q.
\eeq

A possible choice of $Q$ is
\begin{equation}
Q_m \propto T_-^{\beta }P_m ,
\eeq

with $\beta = 2\mu _m +1$. Eigenstates of time operator belong to
different representation.

We have used $SO(2,1)\sim SU(1,1)$ symmetry to solve the problem
of $N$ particles with an inverse-square potential. The conformally
invariant Hamiltonian has a continuous spectrum and is bounded
from below, but its lowest eigenstate, zero energy state, is not
normalizable. This is a reflection of the infrared problem present
when there is no scale in the problem.

{\em 3. Infrared problem.} An elegant way to solve this problem
was proposed by de Alfaro, Fubini, and Furlan \cite{aff} when
solving a one-dimensional field theory model. The idea was to
diagonalize a  compact operator, conformally noninvariant, which
in our case is
\begin{equation}
R = \frac{1}{2} (-\frac{1}{\omega }T_+ + \omega T_-)
\eeq

instead of the original Hamiltonian $T_+$. Owing to time
reparametrization, it can be shown that only the generator $R$
could lead to time evolution laws which are acceptable. In the
limit $\omega \rightarrow 0$, we expect a connection with the
$H_{CM}$ problem.

In order to diagonalize $R$, in accordance with \cite{aff} we
introduce appropriate raising and lowering operators
\begin{equation}
L_{\pm } = \frac{1}{2}(\frac{1}{\omega }T_+ + \omega T_-) \pm T_0,
\eeq

which satisfy the $so(2,1)$ algebra:
\begin{equation}
[R,L_{\pm }] = \pm L_{\pm }, \;\;\;\; [L_+ , L_- ] = -2R.
\eeq

In terms of the $L_{\pm }$ operators, we can diagonalize $L_0 = R$
and find the time operator conjugated to $R$. The vacuum states
are now redefined $P_m$ states (8)
\begin{equation}
|0,\mu _m > = e^{\omega T_-}P_m ,
\eeq

such that $L_- |0,\mu _m > = 0$ and $L_0 |0,\mu _m > = - \mu _m
|0,\mu _m >.$

The diagonalization of $L_0$ in terms of the $T_-$ operator is
given by
\begin{equation}
L_0 L_n^{-2\mu _m-1}(2\omega T_-)|0,\mu _m > = (n-\mu
_m)L_n^{-2\mu _m-1}(2\omega T_-)|0,\mu _m >,
\eeq

where $L_n^{\alpha }(\cdot )$ denotes a Laguerre polynomial. In
terms of $L_+$, solutions are
\begin{equation}
L_0 L_+^n |0,\mu _m > = (n - \mu _m)L_+^n |0,\mu _m >.
\eeq

The time operator canonically conjugated to $L_0 = R$ can be
determined from the equation
\begin{equation}
[L_0, f(L_{\pm })] = \pm L_{\pm }f'(L_{\pm }).
 \eeq

We find
\begin{equation}
\hat{T}_R = c_+ ln L_+ + c_- ln L_- + g(L_0),
\eeq

where $g(L_0)$ is an arbitrary function of $L_0$, and $c_+ - c_- =
1$.

{\em Conclusion.} The property of conformal invariance is used to
construct the time operator in terms of the generators of the
conformal group. For an $N$-particle system the generators of the
conformal group are collective variables, emphasizing the overall
properties of the system. The time operator originated when we
constructed the algebraic solution of the Schroedinger equation.
This is represented by the Barut-Girardello coherent state which
diagonalized the Hamiltonian. The time operator plays the role of
a creation operator for the excitations and is the energy-shift
operator in the system. For large distances $x_i$ or large momenta
$p_i$, from (15) we can show that the time operator assumes the
form of a generalized collective "time-of-arrival" operator,
which, for the $N=2$ case, coincides with the known case. In spite
of its collective character, the time operator does not commute
with a single-particle coordinate $x_i$ or with  particle momenta
$p_i$, and represents a nontrivial case of time noncommutative
theory.

\section*{ Acknowledgment}

I.A is grateful to Ph. Blanchard and T. Cutright for discussion.
This work was supported by the Ministry of Science and Technology
of the Republic of Croatia.



\begin{thebibliography}{99}
\bibitem{mug}
G. Muga, R. S. Mayato and I. L. Egusquiza, {\em Time in Quantum
Mechanics}, Springer-Verlag, 2002 .
\bibitem{c}
F. Calogero, {\em J. Math. Phys.} {\bf 10}, 2191, 2197 (1969);
{\em J. Math. Phys.} {\bf 12}, 419 (1971).
\bibitem{aff}
V. de Alfaro, S. Fubini and G. Furlan, {\em Nouvo Cimento A} {\bf
34},569(1976).
\bibitem{m}
J. Moser, {\em Adv. Math.} {\bf 16}, 197 (1975).
\bibitem{Pere}
A. M. Perelomov, {\em Generalized Coherent States and Their
Applications}, Springer, Berlin, 1986.
\bibitem{and}
I. Andri\'c  and L. Jonke, \Journal{\prd}{\bf 65}{034707 }{2002},
hep-th/0010033.
\bibitem{Gur}
N. Gurappa P. S. Mohanty and P. K. Panigrahi, Phys. Rev. A {\bf 61
}, 034703 (2000).
\bibitem{bg}
A.O.Barut and L.Girardello, {\em Commun.Math.Phys.} {\bf 21}, 41
(1971).

\end{thebibliography}
\end{document}